\documentclass[prb,preprint]{revtex4-1} 

\usepackage{amsmath} 
\usepackage{amsfonts} 

\usepackage{graphicx}

\begin{document}

\title{The Lorentz-Dirac and Landau-Lifshitz equations from the perspective of modern renormalization theory}

\author{Charles W. Nakhleh}
\affiliation{Sandia National Laboratories, Albuquerque,
NM 87185}
\email{cnakhle@sandia.gov} 
\date{\today}

\begin{abstract}
This paper uses elementary techniques drawn from renormalization theory to derive the Lorentz-Dirac equation for the relativistic classical electron from the Maxwell-Lorentz equations for a classical charged particle coupled to the electromagnetic field. I show that the resulting effective theory, valid for electron motions that change over distances large compared to the classical electron radius, reduces naturally to the Landau-Lifshitz equation. No familiarity with renormalization or quantum field theory is assumed.

\end{abstract}

\maketitle

\section{Introduction}
The theory of a classical charged particle coupled to the electromagnetic field is simple to write down but difficult to solve. The principal difficulty is the divergence of the electromagnetic field on the electron's world-line. Any attempt to properly account for the reaction of the radiation emitted by an accelerated electron on the electron's dynamics must confront this poor behavior. How to do so correctly in classical electrodynamics has been a topic of research for more than a century.~\cite{colemanetal}

My primary objective here is to derive a classical electron equation including radiation reaction in a manner that can be followed by a non-specialist familiar with the rudiments of relativistic classical field theory. I intend to show how renormalization techniques related to those developed in quantum field theory simplify the mathematical labor and cast light on the limits of validity of the resulting equation.

By the end of this paper, you should have a good grasp of the physical concepts that underlie the modern understanding of renormalization theory,~\cite{wilson-nobel} and be well-positioned to enter the (still-active) literature on classical electron theory.

\section{Separating out multiple scales in the Maxwell-Lorentz theory}
The Maxwell-Lorentz theory of the classical electron combines the equation of motion for a relativistic charged particle coupled to the electromagnetic field,~\cite{conventions}
\begin{equation}  
m_0\ddot z^\mu(\tau)  = e_0F^{\mu\nu}(z(\tau))\dot z_\nu (\tau), \label{eq:ptcle}
\end{equation}
with Maxwell's equations,
\begin{equation}  
\partial_\nu F^{\mu\nu}(x) = J^\mu(x).
\end{equation}
Here $x^\mu$ is a general space-time point, $z^\mu(\tau)$ describes the particle world-line, 
\begin{equation}
F^{\mu\nu}(x) = \partial^\mu A^\nu (x) - \partial^\nu A^\mu (x) \label{eq:Fdef}
\end{equation}
defines the electromagnetic field tensor, $F^{\mu\nu}$, in terms of the four-vector potential, $A^\mu$, and
\begin{equation}
J^\mu (x) =e_0\! \int^{\infty}_{-\infty}\dot z^{\mu}(\tau ') \delta^{(4)}(x-z(\tau'))\,d\tau ' \label{eq:current}
\end{equation}
specifies the conserved electron current, $J^\mu$, as a functional of the electron's motion. If we impose the Lorenz condition, in which $\partial_\lambda A^\lambda (x) = 0$, Maxwell's equations simplify to:
\begin{equation} 
-\partial^2 A^\mu (x) = J^\mu(x). \label{eq:maxwell}
\end{equation}

The obvious strategy for deriving an electron equation that accounts for radiation reaction is to solve Eq.~(\ref{eq:maxwell}) with a Green function, evaluate the resulting field on the electron world-line, and then insert that self-field into Eq.~(\ref{eq:ptcle}). Recalling, however, that the electric field of a stationary point charge diverges at the charge's position, we might suspect that a similar divergence problem will crop up here as well. We will soon confirm that suspicion, but for now I would like you to think of the divergence as an indication that the structure of the electromagnetic field near the electron contributes importantly to the electron equation of motion that we seek. Short-distance physics matters.

Continuing our examination of the Maxwell-Lorentz equations, we notice two parameters, $m_0$ and $e_0$, with dimensions of mass and charge. Resist the (natural) temptation to think of these parameters as the physical mass and charge of the electron. Physical parameters have experimental definitions (e.g., the physical mass is the total inertia of the electron in an external field) and until we have analyzed the theory's predictions in these defining contexts, we cannot say how the parameters $m_0$ and $e_0$ are related to the physical mass and charge of the electron. 

Dimensional analysis yields further valuable information. Because we use units in which $c=1$, Eqs.~(\ref{eq:Fdef}) and (\ref{eq:maxwell}) provide the dimensions of the potential and field,
\begin{subequations}   
\begin{align}
[A] &= \frac{Q}{L}, \label{eq:dimA} \\
[F] &= [\partial A] = \frac{Q}{L^2}, \label{eq:dimF}
\end{align}
\end{subequations}
where $Q$ and $L$ denote charge and length dimensions. On combining Eqs.~(\ref{eq:dimA}) and (\ref{eq:dimF}) with Eq.~(\ref{eq:ptcle}), and recalling that $[v]=[dz/d\tau]=1$ and $[a]=L^{-1}$, we discover that $[m_0] = Q^2/L$. Even though we do not yet know how the parameters $e_0$ and $m_0$ are related to the physical electron charge, $e_{\rm phys}$, and mass, $m_{\rm phys}$, we can already see the emergence of a natural length scale in classical electrodynamics, namely, the classical electron radius, $r_c \equiv e^2_{\rm phys} m_{\rm phys}^{-1} (4\pi)^{-1}$.

One final piece of analysis provides some identities that will be used at various places throughout this paper. Contract Eq.~(\ref{eq:ptcle}) with $v_\mu$. Because $F^{\mu \nu}$ is antisymmetric, we find that
\begin{equation}
v\cdot a = v\cdot \frac{dv}{d\tau} = \frac{1}{2} \frac{dv^2}{d\tau} = 0, \label{eq:vdota}
\end{equation}
which implies that $v^2 = {\rm const}$. To ensure that $\tau$ represents the proper time, we choose the constant so that $v^2 = -1$. Applying $d/d\tau$ twice in succession to Eq.~(\ref{eq:vdota}), we see that
\begin{equation}
\dot{a}\cdot v +a^2 = 0,  \label{eq:vdota1}
\end{equation}
and
\begin{equation}
2a\cdot \dot{a}+\dot{a}^2 +v\cdot \ddot{a}=0. \label{eq:vdota2}
\end{equation}
Repeated differentiation yields additional relationships among the electron's higher-order accelerations. 

\subsection{The self-field of the classical electron}
We have good reason to believe that the self-field in Eq.~(\ref{eq:ptcle}) will be problematic. To parametrize just how problematic it might be, we introduce a finite length scale, $\epsilon$. This cutoff, itself arbitrary, is the shortest distance that will be considered in the theory.  In the limit that $\epsilon \to 0$, all length scales are included. Once the cutoff is in place, all computed quantities are finite (and cutoff-dependent). The theory is then said to be ``regularized." Our objective is to make explicit the short-distance structure of the regularized self-field, $F^{\mu\nu}_\epsilon(z)$, by computing the dependence of $F^{\mu\nu}_\epsilon(z)$ on $\epsilon$ as $\epsilon \to 0$.

There are two ways for us to proceed. The straightforward path is to depart from the main text and proceed to the appendices in which I compute, in full detail, the regularized self-field. For a first reading, though, I recommend against this choice. As we will see, we can discover a surprising amount of information about the self-field without detailed computation.

Let us try to construct some possible contributions to the self-field generated by the radiating electron. As the notation indicates, we must construct  $F^{\mu\nu}_\epsilon(z)$ from $\epsilon$, $z^\mu$, and derivatives of $z^\mu$: $v^\mu = \dot{z}^\mu, a^\mu = \dot{v}^\mu, \dot{a}^\mu = \ddot{v}^\mu, \ddot{a}^\mu = \dddot{v}^\mu$, and so forth. Because we know that accelerated charges radiate, we require that first- or higher-order derivatives of $v^\mu$ appear in every term. We also require that the variables entering into $F^{\mu\nu}_\epsilon(z)$ all be evaluated at the same proper time (locality), and, of course, that  $F^{\mu\nu}_\epsilon = - F^{\nu\mu}_\epsilon$. Finally, the dimension of the self-field must be $[F]=Q/L^2$. (The charge dimension will be provided by $e_0$.)

To see how these conditions constrain possible contributions to $F^{\mu\nu}_\epsilon(z)$, consider the local combination $e_0 v^\mu a^\nu$. We must antisymmetrize the Lorentz indices. Because $[e_0v a]=Q/L$, we must divide by some length scale to obtain the correct field dimension. But the only length scale available is the cutoff (because we are dealing with a point particle). These considerations lead us to a self-field contribution
\begin{equation}
F^{\mu\nu}_\epsilon(z)\sim e_0\epsilon^{-1} \left(v^\mu a^\nu - v^\nu a^\mu\right), \label{eq:om1term}
\end{equation} 
which is $O(\epsilon^{-1})$ and therefore sensitive to near-field physics. The generality of this derivation emphasizes that sensitivity to short-distance physics is an intrinsic part of the Maxwell-Lorentz system, not an artifact of any particular solution method.

As another example, consider the product $(d^5 v^\mu/d\tau ^5)( d^8 v^\nu/ d\tau ^8)$. If we impose the correct Lorentz structure, and insert the correct power of $\epsilon$ to get the right dimension, we find an $O(\epsilon^{11})$ self-field contribution 
\begin{equation}
F^{\mu\nu}_\epsilon(z)\sim e_0 \epsilon^{11}\left( \frac{d^5 v^\mu}{d\tau ^5}\frac{d^8 v^\nu}{d\tau ^8} -\frac{d^5 v^\nu}{d\tau ^5}\frac{d^8 v^\mu}{d\tau ^8} \right).
\end{equation} 
Unless the electron's accelerations are extremely violent, this contribution is insensitive to near-field physics.

These two examples point to a systematic method of constructing the self-field expansion:  write down all possible local combinations of the velocity and its derivatives, antisymmetrize appropriately, and adjust the power of $\epsilon$ to make the self-field dimensionally correct. The terms with negative powers of $\epsilon$ are sensitive to near-field physics, those with positive powers of $\epsilon$ are not. Terms that do not depend on $\epsilon$, $O(1)$ terms, are borderline.

This procedure is similar to the multipole expansion from elementary electrostatics. Think of an arbitrary static charge distribution characterized by a length scale, $l$. At distances much greater than $l$ from the distribution, the first few multipole moments adequately represent the potential, which implies that most of the detailed structure of the charge distribution is not relevant to the far field. Furthermore, if the underlying charge distribution obeys any symmetries, the structure of the multipole expansion usually simplifies dramatically.

Keeping this analogy in mind, we can write down the lower-order contributions to the self-field of the classical electron. The only term with a single factor of $\dot{v}=a$ that has the appropriate Lorentz index structure is the $O(\epsilon ^{-1})$ contribution in Eq.~(\ref{eq:om1term}). $O(1)$ terms could be generated in two possible ways: one factor of $\ddot{v}=\dot{a}$ together with a $v$ or two factors of $\dot{v}=a$. Because $a^\mu a^\nu$ vanishes upon antisymmetrization, the sole $O(1)$ term is
\begin{equation}
F^{\mu\nu}_\epsilon(z)\sim e_0 \left(v^\mu \dot{a}^\nu - v^\nu \dot{a}^\mu\right). \label{eq:o1term}
\end{equation} 

More and more terms emerge as we go to higher powers of $\epsilon$, but, up to terms of $O(\epsilon)$ and higher, we have found that
\begin{equation}
F^{\mu\nu}_\epsilon(z) = c_{-1}e_0 \epsilon^{-1} (v^\mu a^\nu - v^\nu a^\mu) +c_0e_0 \left(v^\mu \dot{a}^\nu - v^\nu \dot{a}^\mu\right) +O(\epsilon), \label{eq:self-field} 
\end{equation}
where $\{c_{-1}, c_{0}\}$ are undetermined numerical constants.

Observe the similarity between Eq.~(\ref{eq:self-field}) and the multipole expansion mentioned earlier. In each case, the Taylor expansion lumps the short-distance information into constants that multiply local derivatives of the long-distance degrees of freedom. In each case, we wish to find an approximate description valid at long distances. This limited aim allows us to keep only the first few terms of the expansion.

In the classical electron case, however, to truncate the expansion at a given order in $\epsilon$, say $n$, we must exclude electron motions whose accelerations are so violent that the neglected terms of $O(\epsilon^{n+1})$ become comparable to the retained terms. I will define an electron motion to be ``admissible" if, and only if, the proper-time derivatives of the motion satisfy the restriction $| \epsilon^n d^n v^{\mu}/d\tau^n | \ll 1$, for all $n \ge 1$. Physically, this restriction means that we exclude electron motions that vary over proper-time scales smaller than the cutoff. Admissible motions permit us to safely abridge the self-field expansion. Inadmissible motions offer us no such guarantee. 

\subsection{Normalizing the theory at long distances}
If we insert the self-field from Eq.~(\ref{eq:self-field}) into the particle equation of motion, Eq.~(\ref{eq:ptcle}), and include an external field, we find that
\begin{equation}
\left(m_0- e_0^2c_{-1} \epsilon^{-1}\right) a^\mu= e_0 F^{\mu\nu}_{\rm ext} v_\nu+e_0^2  c_0 \left(\dot{a}^\mu - a^2 v^\mu \right) +O(\epsilon). \label{eq:renorm1}
\end{equation} 

We may now relate the parameters $m_0$ and $e_0$ to the physical mass and charge of the electron. Focus on $m_0$ and consider a slow, very slightly accelerated, motion. In that case, Eq.~(\ref{eq:renorm1}) reduces to:
\begin{equation}
\left(m_0- e_0^2c_{-1} \epsilon^{-1}\right) a^\mu= e_0 F^{\mu\nu}_{\rm ext} v_\nu.
\end{equation} 
The coefficient multiplying the acceleration clearly functions as the electron inertia in the context of this theory. We must choose (``tune") the parameter $m_0$ so that the electron inertia term is the physical mass of the electron:
\begin{equation}
m_0\equiv m_{\rm phys}+e_0^2c_{-1} \epsilon^{-1}. \label{eq:mphys}
\end{equation} 
Notice that the dynamics of the coupled field-particle system allows the short-distance structure (signaled by $\epsilon^{-1}$) of the electromagnetic field to enter into the relationship between the parameter $m_0$ and the physical quantity $m_{\rm phys}$. Notice further that as $\epsilon$ changes, $m_0 = m_0(\epsilon)$ {\it also} changes in such a way that $m_{\rm phys}$ remains {\it unchanged}.

What about $e_0$? Consider an electron at rest at the origin. Put $A^\mu = (\Phi, 0)$. Then Maxwell's equations, Eq.~(\ref{eq:maxwell}), imply that  
\begin{equation}
-\nabla^2\Phi({\bf x}) = e_0 \delta^{(3)}({\bf x}),
\end{equation}
which informs us that the normalization of the charge parameter $e_0$ remains trivial ($e_0 = e_{\rm phys}$) even in the coupled theory. This ``no-charge-renormalization" result is a deep implication of classical electrodynamics (one that does not carry over to quantum electrodynamics).

As an exercise, you should summarize these results in a system of differential equations,
\begin{subequations}
\begin{align}
-\epsilon \frac{dm_0}{d\epsilon}  &= c_{-1}\frac{e_0^2}{\epsilon}, \label{eq:RGeqn1} \\
-\epsilon \frac{de_0}{d\epsilon}  &= 0, \label{eq:RGeqn2}
\end{align}
\end{subequations}
which describes precisely how $m_0$, $e_0$ must vary with the cutoff to leave $m_{\rm phys}$, $e_{\rm phys}$ invariant. The solutions of Eqs.~(\ref{eq:RGeqn1}) and~(\ref{eq:RGeqn2}) satisfy the condition $\{m_0(\epsilon),e_0(\epsilon)\} \to \{m_{\rm phys},e_{\rm phys}\}$ as $\epsilon \to \infty$. Such ``renormalization group" equations play an important role in more advanced treatments of renormalization.

If we look back over our work, we see that we have managed to extract a great deal of physical information out of the Maxwell-Lorentz theory with few detailed computations. The sensitivity of the mass parameter to short-distance physics has been brought out, and the form of the reaction force on the radiating electron has been arrived at quite simply. Moreover, we have found that we must restrict the electron motions that we can allow into our effective description of the radiating electron.

To proceed further, we require the detailed computations in the appendices to extract the constants in the self-field expansion. For the regulator used in this paper,
\begin{equation}
c_{-1} = -\frac{1}{2}\left( \frac{1}{4\pi} \right), \label{eq:c1}
\end{equation}
while
\begin{equation}
c_{0} = \frac{2}{3}\left( \frac{1}{4\pi} \right), \label{eq:c0}
\end{equation}
regardless of the method of regularization. If we insert Eqs.~(\ref{eq:mphys}), (\ref{eq:c1}), and (\ref{eq:c0}) into Eq.~(\ref{eq:renorm1}) we arrive at the standard Lorentz-Dirac equation:~\cite{dirac1938}
\begin{equation}
m_{\rm phys} a^\mu(\tau)= e_{\rm phys} F^{\mu\nu}_{\rm ext} v_\nu+\frac{2}{3}\left(\frac{e_{\rm phys}^2}{4 \pi} \right) \left(\dot{a}^\mu - a^2 v^\mu \right) +O(\epsilon),\label{eq:ldeqn}
\end{equation} 
where the ``$O(\epsilon)$" reminds us that we are dropping higher order terms---dubbed ``structure" terms in the older literature---in the effective-interaction expansion of the self-field. Remember that this equation applies only to {\it admissible} electron motions. For inadmissible electron motions, the structure terms are no longer negligible, and the Lorentz-Dirac equation is no longer valid.

\section{Enforcing consistency and eliminating runaways}
The restriction to admissible electron motions has important ramifications for classical electron theory. Because the natural scale in classical electrodynamics is the classical electron radius, take $\epsilon \sim r_c$ in what follows. Following tradition, I define $\epsilon = 2r_c/3 \equiv \tau_0$. With this choice, Eq.~(\ref{eq:ldeqn}) transforms into\cite{endnote5}
\begin{equation}
a^\mu(\tau)= f^\mu(\tau)+\tau_0 \left(\dot{a}^\mu - a^2 v^\mu \right) +O(\tau_0^2),\label{eq:ldeqn2}
\end{equation} 
where $f^\mu \equiv e_{\rm phys} F^{\mu\nu}_{\rm ext} v_\nu /m_{\rm phys}$. 

Equation (\ref{eq:ldeqn2}) presents some puzzling questions that date back to the earliest investigations in this area. The source of the trouble is the $\dot{a}^\mu$ term. If we regard Eq.~(\ref{eq:ldeqn2}) as an initial-value problem, we have to specify the electron's initial {\it acceleration}, in addition to the usual position and velocity. Contracting Eq.~(\ref{eq:ldeqn2}) with $a_\mu$, and neglecting the external field, we find that
\begin{equation}
a^2 = \tau_0 \left( a\cdot \dot{a} - a^2 v\cdot a \right)=\tau_0 a\cdot \dot{a} = \frac{\tau_0}{2} \frac{d a^2}{d\tau},
\end{equation}
which has an exponentially increasing solution, $a^2 \propto \exp{\left(2\tau/\tau_0\right)}$, in addition to the expected solution, $a^\mu = 0$, (or $v^\mu={\rm const}$). This additional solution is aptly dubbed a ``runaway." Had we chosen to work to a higher order in $\tau_0$, even higher derivatives would have appeared in the equations, necessitating the initial specification of the third, fourth, $\ldots$, derivatives of the electron position, leading to additional runaway solutions. Observe that the derivative expansion of the self-field is the source of runaways.

The motion described by the runaway is blatantly inadmissible. Because we derived the Lorentz-Dirac equation as a long-distance effective theory, we must ensure that its solutions do not violate the assumptions that our derivation presupposed. In other words, runaways must be excluded.

In a very brief note,\cite{bhabha} Bhabha pointed out that runaways are not analytic functions of $\tau_0$ near $\tau_0 = 0$. He proposed discarding all such solutions as unphysical. His criterion turns out to be closely related to our work. We discovered that we could ensure convergence of the long-distance expansion of the self-field only if we restricted our attention to admissible electron motions, which automatically satisfy Bhabha's analyticity criterion (because they are---by definition---Taylor-expandable in the short-distance cutoff, $\tau_0$).

This train of thought leads naturally to the idea that simple perturbation theory in the cutoff parameter could eliminate runaway solutions to the Lorentz-Dirac equation. This basic idea has been generalized to a wide variety of effective theories that appear in such diverse contexts as quantum field theory, string theory, and general relativity. All these effective long-distance theories involve high-order derivatives and runaway solutions, and all require some sort of constraint to eliminate the unphysical solutions.\cite{simon1990}

The method of perturbative constraints is straightforward to implement for classical electron theory. We use Eq.~(\ref{eq:vdota1}) to rewrite Eq.~(\ref{eq:ldeqn2}) as
\begin{equation}
a^\mu(\tau)= f^\mu(\tau)+\tau_0\Pi^{\mu\nu}\dot{a}_\nu +O(\tau_0^2), \label{eq:ldeqn3}
\end{equation}
where $\Pi^{\mu \nu}= g^{\mu\nu} +v^\mu v^\nu$ is a projector that satisfies $\Pi^{\mu \nu}v_\nu = 0$. We now seek a perturbative solution to Eq.~(\ref{eq:ldeqn3}) in the small parameter $\tau_0$. If we insert the expansion $a^\mu = \underset{0}{a}^\mu+\tau_0  \underset{1}{a}^\mu+O(\tau_0^2)$ into Eq.~(\ref{eq:ldeqn3}), we find that
\begin{equation}
\underset{0}{a}^\mu+\tau_0  \underset{1}{a}^\mu+O(\tau_0^2)=  f^\mu(\tau)+\tau_0\underset{0}{\Pi}^{\mu \nu} \underset{0}{\dot{a}_\nu} +O(\tau_0^2).
\end{equation}
Now match powers of $\tau_0$ to find the $O(1)$ equation, $\underset{0}{a}^\mu=f^\mu$, and the $O(\tau_0)$ equation,
\begin{equation}
\underset{1}{a}^\mu = \underset{0}{\Pi}^{\mu \nu} \underset{0}{\dot{a}_\nu} =  \underset{0}{\Pi}^{\mu \nu} \dot{f}_\nu,
\end{equation}
into which we have inserted the result of the $O(1)$ equation. The perturbatively constrained Lorentz-Dirac equation is then:
\begin{equation}
a^\mu = f^\mu + \tau_0 \Pi^{\mu \nu} \dot{f}_\nu +O(\tau_0^2), \label{eq:spohn}
\end{equation}
where we have replaced $\underset{0}{\Pi}^{\mu \nu}$ with $\Pi^{\mu \nu}$, a substitution permitted to the order of approximation indicated. For the expansion in $\tau_0$ to remain well-behaved, the external force must satisfy the condition $| \tau_0  \Pi^{\mu \nu} \dot{f}_\nu | \ll |f^\mu |$. External forces satisfying this condition will also be called ``admissible."

Note that the enforcement of the perturbative constraint has eliminated the $\dot{a}$ term. And if we consider the case of vanishing external force, for which Eq.~(\ref{eq:spohn}) simplifies to $a^\mu = 0$, with solution $v^\mu={\rm const}$, we see that the runaway solution has been eliminated as well. 

\section{Conclusions}
Rohrlich\cite{rohrlich2008} calls Eq.~(\ref{eq:spohn}) the ``physical Lorentz-Abraham-Dirac" equation. It is also often called the ``Landau-Lifshitz" equation. \cite{landaulifshitz} Vigorous debates continue in the literature\cite{griffiths2009} about the relationship of this equation to the Lorentz-Dirac equation and their respective domains of validity. Our work throws some light on these questions. I have argued that deriving an effective description of the radiating electron from the regularized Maxwell-Lorentz theory leads directly to the Lorentz-Dirac equation. This description is valid for electron motions---admissible motions---that vary only over proper-time scales large compared to the classical electron radius. When we enforce the constraint of admissibility explicitly on the Lorentz-Dirac equation, the Landau-Lifshitz equation---together with the condition of validity that the external force be admissible---emerges naturally. 

\appendix
\section{Regularization and explicit computation of the self-field}
There is no royal road to renormalization. Explicit computation of the self-field is both essential and informative. We integrate Eq.~(\ref{eq:maxwell}) using the retarded Green function:\cite{colemanetal}
\begin{equation}
D_R(x) = \frac{1}{2\pi} \theta(t) \delta(x^2),
\end{equation}
where $x^\mu=(t,{\bf x})$ and $x^2=|{\bf x}|^2-t^2$. Up to a free field (the ``in" field), which is taken to vanish,\cite{endnote6} the solution for the retarded potential at a general space-time point $x^\mu$ is:
\begin{equation}
A^\mu_R (x) = e \int_{-\infty}^{\infty} d\tau \, D_R(x-z(\tau)) v^\mu (\tau) = \frac{e}{2\pi} \int_{-\infty}^{0} d\tau \delta((x-z(\tau))^2)v^\mu (\tau), \label{eq:retpot}
\end{equation}
where the $\theta$-function constraint has been taken into account in the latter term, and we have defined the origin of proper time by  $z^0(\tau=0)=t=x^0$. (We have put $e_{\rm phys} = e$ for simplicity.) As long as the point $x^\mu$ is off the particle world-line, this formula is well-defined and yields the usual Li\'{e}nard-Wiechert potentials.

What is needed in Eq.~(\ref{eq:ptcle}) is the value of the field at a point on the world-line, say $x^\mu=z^\mu(0)$. In that case, Eq.~(\ref{eq:retpot}) diverges and requires regularization. The convenient regulator we will use replaces Eq.~(\ref{eq:retpot}) with:\cite{galtsov}
\begin{equation}
A^\mu_\epsilon (x) = \frac{e}{2\pi} \int_{-\infty}^{0} d\tau \delta((x-z(\tau))^2+\epsilon^2)v^\mu (\tau). \label{eq:retpot2}
\end{equation}
Although we wish ultimately to compute the field, a useful warm-up exercise is to compute the retarded potential on the electron world-line. We can use the expansions
\begin{subequations}
\begin{align}
z^\mu(\pm\tau) &= z^\mu(0) \pm v^\mu(0) \tau + a^\mu(0) \tau^2/2 + O(\tau^3), \\
\left(z(\pm\tau)-z(0)\right)^2 &= v(0)^2\tau^2 + O(\tau^3) = -\tau^2 +O(\tau^3),
\end{align}
\end{subequations}
in Eq.~(\ref{eq:retpot2}) to express the delta function as (recall $x^\mu=z^\mu(0)$):
\begin{equation}
\delta(-\tau^2+\epsilon^2) = \delta(\tau^2-\epsilon^2)=\frac{1}{2\epsilon} \left(\delta(\tau-\epsilon)+\delta(\tau+\epsilon)\right), \label{eq:delta1}
\end{equation}
and immediately compute the potential,
\begin{equation}
A^\mu_\epsilon (z(0)) = \frac{e}{4\pi} \frac{v^\mu (-\epsilon)}{\epsilon}.
\end{equation}
Expanding $v^\mu(-\epsilon)=v^\mu(0)-\epsilon a^\mu(0) +O(\epsilon^2)$, we find that
\begin{equation}
A^\mu_\epsilon (z(0)) = \frac{e}{4\pi} \left( \frac{v^\mu (0)}{\epsilon}-a^\mu(0)\right)+O(\epsilon),
\end{equation}
which can be generalized to any point on the particle world-line:
\begin{equation}
A^\mu_\epsilon (z(\tau)) = \frac{e}{4\pi} \left( \frac{v^\mu (\tau)}{\epsilon}-a^\mu(\tau)\right)+O(\epsilon). \label{eq:cutpot}
\end{equation}
Note that the divergent term has the structure of a relativistic Coulomb potential and is present even for an unaccelerated motion, in complete accord with our intuitive notions of the short-distance coupling of the near-field to the electron. The simultaneous presence of divergent and finite terms in Eq.~(\ref{eq:cutpot}) is a clear sign of the multiscale nature of the potential.

The computation of the field is more involved. Take a derivative of the potential with respect to $x$ inside the integral to obtain:
\begin{equation}
\partial^\lambda A^\mu_\epsilon (x) = \frac{e}{2\pi} \int_{-\infty}^{0} d\tau \delta'((x-z(\tau))^2+\epsilon^2)2\left(x-z(\tau)\right)^\lambda v^\mu (\tau).
\end{equation}
Then insert a very useful identity from Dirac's 1938 paper:~\cite{dirac1938}
\begin{equation}
\delta'((x-z(\tau))^2+\epsilon^2) = - \frac{1}{2} \frac{1}{(x-z(\tau))\cdot v(\tau)}\frac{d}{d\tau} \delta((x-z(\tau))^2+\epsilon^2), \label{eq:dirac}
\end{equation}
and integrate by parts to obtain the compact form:
\begin{equation}
\partial^\lambda A^\mu_\epsilon (x) = \frac{e}{2\pi} \int_{-\infty}^{0} d\tau \delta((x-z(\tau))^2+\epsilon^2)\frac{dL^{\lambda\mu}(\tau)}{d\tau},\label{eq:regA}
\end{equation}
where the important quantity $L^{\lambda\mu}(\tau)$ is defined by:
\begin{equation}
L^{\lambda\mu}(\tau) \equiv \frac{\left(x-z(\tau)\right)^\lambda v^\mu (\tau)}{\left(x-z(\tau)\right)\cdot v(\tau)}. \label{eq:ltensor}
\end{equation}
Defining $N^{\lambda\mu}(\tau)\equiv dL^{\lambda\mu}(\tau)/d\tau$, we can use Eq.~(\ref{eq:delta1})  to integrate Eq.~(\ref{eq:regA}) to
\begin{equation}
\partial^\lambda A^\mu_\epsilon (x) = \frac{e}{4\pi}\frac{N^{\lambda\mu}(-\epsilon)}{\epsilon}, \label{eq:regAfinal}
\end{equation}
from which we compute the antisymmetric combination:
\begin{equation}
F^{\lambda \mu}_\epsilon (x)= \frac{e}{4\pi}\frac{N^{\lambda\mu}(-\epsilon)-N^{\mu\lambda}(-\epsilon)}{\epsilon}\equiv \frac{e}{4\pi} \frac{N^{[\lambda\mu]}(-\epsilon)}{\epsilon}. \label{eq:regF}
\end{equation}
Now we use Eqs.~(\ref{eq:N1})--(\ref{eq:N3}) to express the regularized self-field up to $O(\epsilon)$:
\begin{subequations}
\begin{align}
F^{\lambda \mu}_\epsilon (z)=&-\frac{1}{2\epsilon} \left(\frac{e}{4\pi}\right) \left(v^\lambda a^\mu - v^\mu a^\lambda \right) \\
&+\frac{2}{3}\left(\frac{e}{4\pi}\right)  \left(v^\lambda \dot{a}^\mu - v^\mu \dot{a}^\lambda \right) \\
&-{3\epsilon\over 8} \left(\frac{e}{4\pi}\right) \left[ \left(v^\lambda \ddot{a}^\mu -v^\mu\ddot{a}^\lambda\right)+{2\over 3} \left(a^\lambda \dot{a}^\mu -a^\mu \dot{a}^\lambda\right)+ {2 v\cdot \dot{a} \over 3} \left(v^\lambda a^\mu-v^\mu a^\lambda \right)  \right]. \label{eq:self-field-eps}
\end{align}
\end{subequations}
This explicit computation justifies the constants given in Eqs.~(\ref{eq:c1}) and (\ref{eq:c0}). I have included the $O(\epsilon)$ contribution to the self-field to show you the structure of the self-field expansion and help you practice these types of manipulations.

A subtle consistency check on our results comes from computing $dA^\mu / d\tau$ in two different ways. First, applying $d / d\tau$ directly to Eq.~(\ref{eq:cutpot}), we find that
\begin{equation}
\frac{dA^\mu(z)}{d\tau} =  \frac{e}{4\pi} \left( \frac{a^\mu (\tau)}{\epsilon}-\dot{a}^\mu(\tau)\right)+O(\epsilon).  \label{eq:dcutpot}
\end{equation}
On the other hand, the chain rule of differentiation enables us to conclude that
\begin{equation}
 \frac{dA^\mu(z)}{d\tau} = \left( \partial^\lambda A^\mu \right)v_\lambda. \label{eq:dcutpot2}
\end{equation}
Fortunately, using Eqs.~(\ref{eq:N01}) and (\ref{eq:N02}) with Eq.~(\ref{eq:regAfinal}), we find that the two methods yield equivalent results, because
\begin{subequations}
\begin{align}
{N^{\lambda  \mu}(-\epsilon) v_\lambda \over \epsilon }&= -{1\over \epsilon} \left(v^2 a^\mu + \frac{1}{2} v\cdot a v^\mu\right)+\left(v^2 \dot{a}^\mu + v \cdot a a^\mu + \frac{v\cdot \dot{a} v^\mu - a^2 v^2 v^\mu}{3} \right), \\
&= \frac{a^\mu (\tau)}{\epsilon}-\dot{a}^\mu(\tau).
\end{align}
\end{subequations}
Note carefully that this consistency check involves both divergent and finite terms in the regularized expressions.

\section{Expansion of $L^{\lambda \mu}$, $N^{\lambda \mu}$}
Our objective is to expand Eq.~(\ref{eq:ltensor}),
\begin{equation}
L^{\lambda\mu}(\tau) \equiv \frac{\left(z(\tau)-x\right)^\lambda v^\mu (\tau)}{\left(z(\tau)-x\right)\cdot v(\tau)}, \label{eq:Ldef}
\end{equation}
where $z^\mu(\tau=0)=x^\mu$, using the Taylor expansions:
\begin{subequations}
\begin{align}
z^\mu(\tau) &= x^\mu +\tau v_0^\mu +\frac{\tau^2}{2!} a_0^\mu + \frac{\tau^3}{3!} \dot{a}_0^\mu+\frac{\tau^4}{4!} \ddot{a}_0^\mu +O(\tau^5), \\
v^\mu (\tau)&= v_0^\mu +\tau a_0^\mu + \frac{\tau^2}{2!} \dot{a}_0^\mu+\frac{\tau^3}{3!} \ddot{a}_0^\mu +O(\tau^4).
\end{align}
\end{subequations}
Then 
\begin{equation}
\left(z(\tau)-x\right)^\lambda v^\mu (\tau) = \tau A^{\lambda\mu}+\tau^2B^{\lambda\mu}+\tau^3 C^{\lambda\mu} + \tau^4 D^{\lambda\mu} +O(\tau^5), \label{eq:Lnum}
\end{equation}
where
\begin{subequations}
\begin{align}
A^{\lambda\mu}&=v_0^\lambda v_0^\mu, \\
B^{\lambda\mu} &= v_0^\lambda a_0^\mu +{1\over 2} a_0^\lambda v_0^\mu, \\
C^{\lambda\mu} &={1\over 2} v_0^\lambda \dot{a}_0^\mu + {1\over 2} a_0^\lambda a_0^\mu + {1\over 6} \dot{a}_0^\lambda v_0^\mu, \\
D^{\lambda\mu} &= {1\over 6} v_0^\lambda \ddot{a}_0^\mu + {1\over 4} a_0^\lambda \dot{a}_0^\mu +{1\over 6} \dot{a}_0^\lambda a_0^\mu+{1\over 24} \ddot{a}_0^\lambda v_0^\mu,
\end{align}
\end{subequations}
from which it follows that
\begin{equation}
\left(z(\tau)-x\right)\cdot v(\tau) = -\tau +\tau^3 {v_0\cdot \dot{a}_0 \over 6} - \tau^4 {5 \dot{a}_0^2 \over 24} +O(\tau^5).
\end{equation}
If we insert these expressions into Eq.~(\ref{eq:Ldef}), we find that
\begin{subequations}
\begin{align}
L^{\lambda\mu}(\tau) &= -\frac{A^{\lambda\mu}+\tau B^{\lambda\mu}+\tau^2 C^{\lambda\mu} + \tau^3 D^{\lambda\mu} +O(\tau^4)}{1- \tau^2 {v_0\cdot \dot{a}_0 \over 6} + \tau^3 {5 \dot{a}_0^2 \over 24} +O(\tau^4)}, \label{eq:Lnum} \\
&=-\left(1+ \tau^2 {v_0\cdot \dot{a}_0 \over 6} - \tau^3 {5 \dot{a}_0^2 \over 24} +O(\tau^4) \right) \left(A^{\lambda\mu}+\tau B^{\lambda\mu}+\tau^2 C^{\lambda\mu} + \tau^3 D^{\lambda\mu} +O(\tau^4)\right),\\
&=-A^{\lambda\mu}-\tau B^{\lambda\mu} -\tau^2\left( C^{\lambda\mu}+{v_0\cdot\dot{a}_0 \over 6}A^{\lambda\mu}\right)-\tau^3\left( D^{\lambda\mu} + {v_0\cdot\dot{a}_0 \over 6}B^{\lambda\mu}- {5 \dot{a}_0^2 \over 24}A^{\lambda\mu}  \right)+O(\tau^4),
\end{align}
\end{subequations}
which we differentiate to obtain
\begin{equation}
N^{\lambda \mu} (\tau)= -B^{\lambda\mu} -2\tau\left( C^{\lambda\mu}+{v_0\cdot\dot{a}_0 \over 6}A^{\lambda\mu}\right)-3\tau^2\left( D^{\lambda\mu} + {v_0\cdot\dot{a}_0 \over 6}B^{\lambda\mu}- {5 \dot{a}_0^2 \over 24}A^{\lambda\mu}  \right)+O(\tau^3).
\end{equation}
Dropping the ``0" subscripts, we find that
\begin{subequations}
\begin{align}
N^{\lambda \mu} (-\epsilon)&= -\left(v^\lambda a^\mu +{a^\lambda v^\mu \over 2}  \right) \label{eq:N01} \\
&+\epsilon \left[v^\lambda \dot{a}^\mu + a^\lambda a^\mu +{1\over 3} \left( \dot{a}^\lambda v^\mu -a^2 v^\lambda v^\mu\right)  \right] \label{eq:N02} \\
&-3\epsilon^2 \left[ {1\over 6} v^\lambda \ddot{a}^\mu +{1\over 4} a^\lambda \dot{a}^\mu +{1\over 6} \dot{a}^\lambda a^\mu +{1\over 24} \ddot{a}^\lambda v^\mu + {v\cdot \dot{a} \over 6} \left(v^\lambda a^\mu+{a^\lambda v^\mu \over 2} \right) -{5\dot{a}^2 \over 24} v^\lambda v^\mu \right].\label{eq:N03}
\end{align}
\end{subequations}
As an exercise, you should check the Lorenz condition by using Eqs.~(\ref{eq:vdota})--(\ref{eq:vdota2}) to show that $N^\lambda_\lambda (-\epsilon) = 0$ to the order indicated. The last $O(\epsilon^2)$ term in Eq.~(\ref{eq:N03}), which comes from the expansion of the denominator in Eq.~(\ref{eq:Lnum}), is crucial to obtaining this result. We can now compute the antisymmetric sum, 
\begin{subequations}
\begin{align}
N^{[\lambda \mu]} (-\epsilon) &= -{1\over 2}\left(v^\lambda a^\mu -v^\mu a^\lambda \right) \label{eq:N1}\\
&+{2\epsilon\over 3} \left[v^\lambda \dot{a}^\mu -  v^\mu\dot{a}^\lambda \right] \label{eq:N2}\\
&-3\epsilon^2 \left[ {1\over 8} \left(v^\lambda \ddot{a}^\mu -v^\mu\ddot{a}^\lambda\right)+{1\over 12} \left(a^\lambda \dot{a}^\mu -a^\mu \dot{a}^\lambda\right)+ {v\cdot \dot{a} \over 12} \left(v^\lambda a^\mu-v^\mu a^\lambda \right)  \right], \label{eq:N3}
\end{align}
\end{subequations}
which is needed for the self-field computation in Appendix A.

\begin{acknowledgments}
I would like to thank Peter Lepage for first introducing me to the ideas of modern renormalization theory many years ago. I would also like to acknowledge the valuable comments provided by Ryan McBride and an anonymous reviewer. Sandia National Laboratories is a multi-program laboratory managed and operated by Sandia Corporation, a wholly owned subsidiary of Lockheed Martin Corporation, for the U.S. Department of Energy's National Nuclear Security Administration under contract DE-AC04-94AL85000.
\end{acknowledgments}


\begin{thebibliography}{5}
\bibitem{conventions}We use the metric $g^{\mu\nu} = {\rm diag}(-+++)$ and set $c=1=\mu_0=\epsilon_0^{-1}$. The particle coordinate four-vector is $z^\mu (\tau)$, where $\tau$ is the proper time. An overdot indicates differentiation with respect to the proper time: $\dot{z}^\mu \equiv dz^\mu/d\tau \equiv v^\mu$, $\ddot{z}^\mu \equiv a^\mu$, etc. We will often abbreviate inner products, e.g., $a^\lambda b_\lambda \equiv a\cdot b$, $a^\lambda a_\lambda \equiv a^2$. When a space-time point $x^\mu$ is a function argument, the index is dropped, e.g., $f(x)$.
\bibitem{endnote5} The ``$\tau_0^2$" factor in the ``$O$" sign is caused by the $O(\epsilon)$ piece of the self-field, which, when multiplied by $e_{\rm phys}$ and divided by $m_{\rm phys}$, provides a second factor of $\tau_0$ that combines with the explicit factor of $\epsilon \sim \tau_0$.
\end{thebibliography}
\end{document}